\documentstyle[titlepage,12pt]{article}
\begin{document}
\title{Dynamics of cosmic strings and springs;\\a covariant formulation}
\vspace{5 in}
\author{A.L.Larsen\thanks{E-mail: ALLARSEN@nbivax.nbi.dk}\\
Nordita, Blegdamsvej 17, DK-2100 Copenhagen \O, Denmark}
\maketitle
\begin{abstract}
A general family of charge-current carrying string models is investigated. In
the special case of circular configurations in arbitrary axially symmetric
gravitational and electromagnetic backgrounds the dynamics is determined
by simple point particle Hamiltonians.
A certain "duality" transformation relates our results to previous ones,
obtained by Carter et.al., for an infinitely long open stationary string
in an arbitrary stationary background.
\end{abstract}
\newpage
Classical circular string configurations, obtained for instance as solutions
to the equations of motion for the Nambu-Goto action [1], are known to collapse
unless there are internal or external forces balancing the string tension.
Such forces may arise from charges and currents on the string [2] and/or from
external gravitational and electromagnetic potentials [3-6]. For suitable
internal
or external forces stable non-collapsing configurations may exist, which are
then sometimes denoted as "cosmic springs".
\vskip 6pt
Internal electromagnetic degrees of freedom as well as the coupling to external
potentials can be introduced in different ways by generalizing the ordinary
Nambu-Goto action. One possibility is a Kaluza-Klein construction [4]
involving one or several extra dimensions. Another possibility, originally due
to Witten [5], is to start from a suitable underlying field theory. After
integrating over the transverse dimensions of a linear topological defect
solution, an effective
world-sheet action can be obtained [6].

In this paper we will however consider more general string actions [7]
having the above mentioned models as special examples.
\vskip 6pt
The most important aspect concerning circular strings is the question of
stability. A convenient way to deal with this question is to consider string
loops with time-dependent radius. A Hamiltonian formulation with an effective
potential for the string radius will then in general make it very easy to
separate the stable configurations from the unstable (collapsing) ones.
An important aim of this paper is therefore to develop a Hamiltonian
formalism for circular string configurations obtained as solutions in very
general string models. For completeness we will also allow the string to
propagate in the vertical direction, perpendicular to the plane of the
string. It will then turn out that the circular string is described by the
same coordinates as an ordinary (charged) point particle in an axially
symmetric gravitational and electromagnetic background. Mathematical questions
such as Hamilton-Jacobi separability of the variables in special backgrounds
arise naturally in this context, just as in the point particle case.
\vskip 6pt
The non-linearity of the equations of motion for a string in a non-trivial
background, makes it extremely difficult to analyze the dynamics unless a
suitable {\it Ansatz} is made, thus reducing the amount of freedom. A very
popular class of configurations studied in the literature, besides the
class of circular strings, is represented
by infinitely long open stationary strings in stationary backgrounds [8,9].
As a curiosity we find that there is a very simple "duality" transformation
relating these 2 categories of solutions.
\vskip 6pt
The starting point of our analysis will be a charge-current carrying string
described by the following action [7]:
\begin{equation}
{\cal S}=\int L\sqrt{-\det G_{\alpha\beta}} d\tau d\sigma,
\end{equation}
where $G_{\alpha\beta}$ is the induced metric on the world-sheet:
\begin{equation}
G_{\alpha\beta}=g_{\mu\nu}x^\mu_{,\alpha}x^\nu_{,\beta},
\end{equation}
and $L$ is the Lagrangian density (being just a constant for the ordinary
Nambu-Goto string [1]). $x^\mu$ $(\mu=0,1,2,3)$ are the spacetime coordinates
and
$g_{\mu\nu}$ is the metric of curved 4-dimensional spacetime.

A very general family of string models is obtained by
letting the Lagrangian density $L$ be a function of the world-sheet projection
of
the gauge covariant derivative
of a world-sheet scalar field $\Phi$:
\begin{equation}
L=L(\omega);\hspace*{5 mm}\omega\equiv G^{\alpha\beta}(\Phi_{,\alpha}+
A_\mu x^\mu_{,\alpha})(\Phi_{,\beta}+A_\mu x^\mu_{,\beta}),
\end{equation}
where $A_\mu$ is the external electromagnetic potential. We are interested in
circular strings and will therefore restrict ourselves by considering only
axially symmetric gravitational and electromagnetic backgrounds. It is then
convenient to decompose the spacetime coordinates in the following way:
\begin{equation}
x^\mu=(x^a,\phi);\hspace*{5 mm}a=0,1,2,
\end{equation}
where $\phi$ is a periodic coordinate that can be identified with the
string-parameter $\sigma$ of the circular string. An axially symmetric
background is obtained by considering only metrics and electromagnetic
potentials independent of $\phi$. Except for this restriction $g_{\mu\nu}$ and
$A_\mu$ will be arbitrary in the following.

The {\it Ansatz} for our circular string is besides $\phi=\sigma$ taken to be:
\begin{equation}
x^a=x^a (\tau),\hspace*{5 mm}\Phi=f(\tau)+n\sigma,
\end{equation}
where $f$ is an arbitrary function of $\tau$ and $n$ is a constant. Thinking
for instance in terms of polar spherical coordinates in 3-dimensional space
we can identify $x^a=(
t,r,\theta)$. The spacetime dynamics of the string configurations described by
the {\it
Ansatz} (5) then consists of vertical propagation $\theta(\tau)$,
perpendicular to the plane
of the string, combined with radial "oscillations" $r(\tau)$.

Using the {\it Ansatz} (5) we find the components of the induced metric on
the world-sheet:
\begin{equation}
G_{00}=g_{ab}\dot{x}^a\dot{x}^b,\hspace*{5 mm}G_{11}=g_{33}\equiv g_{\phi\phi},
\hspace*{5 mm}G_{01}=G_{10}=g_{a3}\dot{x}^a\equiv g_{a\phi}\dot{x}^a,
\end{equation}
where a dot denotes derivative with respect to $\tau$. It follows that:
\begin{equation}
\det G_{\alpha\beta}=(g_{ab}g_{\phi\phi}-g_{a\phi}g_{b\phi})\dot{x}^a \dot{x}^b
\equiv\gamma_{ab}\dot{x}^a\dot{x}^b\equiv \gamma,
\end{equation}
where $\gamma_{ab}$ up to a rescaling is the 3-dimensional quotient spacetime
metric:
\begin{equation}
\gamma_{ab}=g_{\phi\phi}(g_{ab}-\frac{g_{a\phi}g_{b\phi}}{g_{\phi\phi}}).
\end{equation}
{}From (3),(5) and (6) we also find:
\begin{equation}
\omega=\frac{1}{\gamma g_{\phi\phi}}\left(g_{\phi\phi}(\dot{f}+A_a\dot{x}^a)-
g_{a\phi}\dot{x}^a (n+A_\phi)\right) ^2+\frac{(n+A_\phi)^2}{g_{\phi\phi}}.
\end{equation}
An important point is now that although the {\it Ansatz} (5) has a
non-trivial $\sigma$-dependence, this $\sigma$-dependence drops out of the
action (1).
This follows from the fact that both $\phi$ and $\Phi$ are cyclic coordinates:
The cyclicity of $\phi$ is due to the restriction to consider only axially
symmetric
backgrounds whereas the cyclicity of $\Phi$ follows from the construction of
$L=L(\omega)$ (3). The dynamics of our circular string is therefore fully
determined by the effective action:
\begin{equation}
{\it S}_{eff}=\int \sqrt{-\gamma_{ab}\dot{x}^a\dot{x}^b}L(\omega (\tau))d\tau.
\end{equation}

We now consider the equations of motion for this effective action. The equation
of motion for the function $f(\tau)$ introduced in (5) becomes:
\begin{equation}
\dot{f}+A_a\dot{x}^a=\frac{1}{g_{\phi\phi}}\left(\frac{\Omega\sqrt{-\gamma}}{
2dL/d\omega}+g_{a\phi}\dot{x}^a (n+A_\phi)\right),
\end{equation}
where $\Omega$ is an integration constant. Using this equation we can rewrite
equation (9) as:
\begin{equation}
\omega (\tau)=-\frac{1}{g_{\phi\phi}}\left(\frac{\Omega^2}{4(dL/d\omega)^2}-
(n+A_\phi)^2\right).
\end{equation}
The equations of motion for $x^a$ are somewhat more complicated. Using equation
(11) we get:
\begin{eqnarray}
\frac{d}{d\tau}\hspace*{-2mm}&[&\hspace*{-2mm}\Omega(-\frac{g_{a\phi}}{
g_{\phi\phi}}(n+A_\phi)+A_a)+\frac{\gamma_{ab}}{\sqrt{-\gamma}}(L(\omega)+\frac{
\Omega^2}{2g_{\phi\phi}dL/d\omega})\dot{x}^b]\nonumber\\
\hspace*{-2mm}&-&\hspace*{-2mm}\Omega(-\frac{g_{b\phi}}{g_{\phi\phi}}(n+A_\phi)
+A_b)_{,a}\dot{x}^b+\sqrt{-\gamma}\frac{dL}{d\omega}(\frac{(n+A_\phi)^2}{
g_{\phi\phi}})_{,a}\nonumber\\
\hspace*{-2mm}&-&\hspace*{-2mm}\frac{\Omega^2\sqrt{-\gamma}}{4g^2_{\phi\phi}
dL/d\omega}g_{\phi\phi,a}-\frac{\gamma_{bc,a}\dot{x}^b\dot{x}^c}{2\sqrt{
-\gamma}}(L(\omega)+\frac{\Omega^2}{2g_{\phi\phi}dL/d\omega})=0.
\end{eqnarray}
It is now convenient to introduce the rescaled metric $H_{ab}$:
\begin{equation}
H_{ab}\equiv\gamma_{ab}(L(\omega)+\frac{\Omega^2}{2g_{\phi\phi}dL/d\omega})^2,
\end{equation}
the "quotient spacetime electromagnetic potential" ${\cal A}_a$:
\begin{equation}
{\cal A}_a\equiv-\frac{g_{a\phi}}{g_{\phi\phi}}(n+A_\phi)+A_a,
\end{equation}
as well as the new string time parameter $\tilde{\tau}$:
\begin{equation}
d\tilde{\tau}\equiv\sqrt{-\gamma}(L(\omega)+\frac{\Omega^2}{2g_{\phi\phi}dL/d\omega})
d\tau.
\end{equation}
After some algebra we then obtain the equations of motion (13) in the simple
form:
\begin{equation}
\frac{d^2 x^a}{d\tilde{\tau}^2}+\Gamma^a_{bc}\frac{d x^b}{d\tilde{\tau}}
\frac{d x^c}{d\tilde{\tau}}=\Omega{\cal F}^a\hspace*{1mm}_b\frac{d x^b}{
d\tilde{\tau}},
\end{equation}
where $\Gamma^a_{bc}$ is the Christoffel symbol for the metric $H_{ab}$ whereas
${\cal F}_{ab}$ is the field strenght of the potential ${\cal A}_a$:
\begin{equation}
{\cal F}_{ab}\equiv{\cal A}_{b,a}-{\cal A}_{a,b}.
\end{equation}
The constant Hamiltonian corresponding to the equations of motion (17) is
finally
given by:
\begin{equation}
{\cal H}=\frac{1}{2}H^{ab}(P_a-\Omega{\cal A}_a)(P_b-\Omega{\cal A}_b)+
\frac{1}{2}=0.
\end{equation}
Thus we have shown that the dynamics of the circular string is
determined by the usual equations/Hamiltonian (17),(19) of a point particle of
charge $\Omega$ propagating in a certain "unphysical" curved 3-dimensional
spacetime
$H_{ab}$ and a certain "unphysical" electromagnetic potential ${\cal A}_a$.

Mathematically this seems to be the most elegant formulation of the circular
string configuration for models described by action integrals of the form (1).
It is however desirable to have a simple description involving the physical
spacetime metric $g_{\mu\nu}$ and the physical electromagnetic
potential $A_\mu$
also. First note that the contravariant metric $H^{ab}$ of equation (19) is
given by:
\begin{equation}
H^{ab}=\frac{g^{ab}}{g_{\phi\phi}}(L(\omega)+\frac{\Omega^2}
{2g_{\phi\phi}dL/d\omega})^{-2},\hspace*{5mm}H^{ab}H_{bc}=\delta^a_c,
\end{equation}
where $g^{ab}$ represents the $(ab)$ component of $g^{\mu\nu}$ (thus $g^{ab}
g_{bc}\neq\delta^a_c$ !). We can then perform a conformal rescaling of the
Hamiltonian (19):
\begin{equation}
{\cal H}\longrightarrow\hat{{\cal H}}\equiv g_{\phi\phi}(L(\omega)+\frac{
\Omega^2}{2g_{\phi\phi}dL/d\omega})^2 {\cal H},
\end{equation}
and a corresponding redefinition of the string time parameter:
\begin{equation}
(L(\omega)+\frac{\Omega^2}{2g_{\phi\phi}dL/d\omega})^{-2}\frac{d\tilde{\tau}}{
g_{\phi\phi}}\equiv d\hat{\tau}.
\end{equation}
The new Hamiltonian is:
\begin{equation}
\hat{{\cal H}}=\frac{1}{2}g^{ab}(P_a-\Omega{\cal A}_a)(P_b-\Omega{\cal A}_b)+
\frac{1}{2}g_{\phi\phi}(L(\omega)+\frac{\Omega^2}{2g_{\phi\phi}dL/d\omega})^2=0.
\end{equation}
Finally this Hamiltonian can be written in the 4-dimensional physical covariant
form:
\begin{equation}
\hat{{\cal H}}=\frac{1}{2}g^{\mu\nu}(P_\mu-\Omega A_\mu)(P_\nu-\Omega A_\nu)+
\frac{1}{2}g_{\phi\phi}(L(\omega)+\frac{\Omega^2}{2g_{\phi\phi}dL/d\omega})^2-
\frac{\Omega^2}{2}\frac{(n+A_\phi)^2}{g_{\phi\phi}}=0,
\end{equation}
with the extra constraint on the conserved angular momentum of the
circular string:
\begin{equation}
P_\phi=-n\Omega.
\end{equation}
The equivalence of the 2 Hamiltonians (23) and (24) can easily be established
by
comparing the corresponding Hamilton equations. In the form (24) the circular
string is described by the usual Hamiltonian of a point particle of charge
$\Omega$ propagating in the 4-dimensional physical spacetime $g_{\mu\nu}$,
the physical
electromagnetic potential $A_\mu$ and in the extraordinary scalar potential:
\begin{equation}
V\equiv\frac{1}{2}g_{\phi\phi}(L(\omega)+\frac{\Omega^2}{2g_{\phi\phi}
dL/d\omega})^2-\frac{\Omega^2}{2}\frac{(n+A_\phi)^2}{g_{\phi\phi}}.
\end{equation}
Until now the Lagrangian density $L(\omega)$ of the action (1) has been treated
as a completely arbitrary function of $\omega$. To obtain the potential (26) as
an explicit function of $x^a$ the procedure is to solve equation (12) for
$\omega(x^a(\tau))$, and then to write $L$ and $dL/d\omega$, appearing in (26),
in terms of $x^a(\tau)$. Solving equation (12) analytically with respect to
$\omega$ is however not possible in general. Fortunately it is possible in the
2 most popular models studied in the literature and in both cases relatively
simple expressions for the potential (26) are obtained:

The Kaluza-Klein model originally developed by Nielsen [4] is in the formalism
of this paper described by the Lagrangian density $L(\omega)=\sqrt{1+\omega}$.
In this case equation (12) is solved by:
\begin{equation}
\omega=\frac{(n+A_\phi)^2-\Omega^2}{g_{\phi\phi}+\Omega^2},
\end{equation}
and the potential (26) leads to:
\begin{equation}
V=\frac{1}{2}(g_{\phi\phi}+\Omega^2+(n+A_\phi)^2),
\end{equation}
in agreement with Ref.10.

Another popular model [6] originally developed by Witten [5] is obtained by the
choice $L(\omega)=1+\omega/2$. In that case equation (12) leads to:
\begin{equation}
\omega=\frac{(n+A_\phi)^2-\Omega^2}{g_{\phi\phi}},
\end{equation}
with the following expression for the potential (26):
\begin{equation}
V=\frac{1}{2}(g_{\phi\phi}+\Omega^2+(n+A_\phi)^2+\frac{(\Omega^2-
(n+A_\phi)^2)^2}{8g_{\phi\phi}},
\end{equation}
also in agreement with Ref.10, where an action integral adjoint to (1) was
used.
\vskip 6pt
The derivation of the Hamiltonians (19) and (24) is very similar to the
derivation of the corresponding Hamiltonians for
an infinitely long open stationary string in a stationary background,
carried out by Carter et.al. [9]. The essential difference (except for
differences
in notation etc.) seems to be a formal interchange of the azimuthal angle
$\phi$
and the coordinate time $t$. Looking backwards we can understand this
similarity
in the following way: Decomposing the coordinates $x^a$ of equation (4) in the
way:
\begin{equation}
x^a=(t,x^i);\hspace*{5mm}i=1,2,
\end{equation}
where $t$ is the time coordinate, and then formally performing the
interchangings:
\begin{equation}
\tau\longleftrightarrow\sigma,\hspace*{1cm}t\longleftrightarrow\phi
\end{equation}
in the {\it Ansatz} (5), we get the "dual" {\it Ansatz}:
\begin{eqnarray}
t=\tau,\hspace*{5mm}\phi=\phi(\sigma)\hspace*{-3mm}&,&\hspace*{3mm}
x^i=x^i (\sigma);\hspace*{2mm}i=1,2\nonumber\\
\Phi\hspace*{-2mm}&=&\hspace*{-2mm}f(\sigma)+n\tau.
\end{eqnarray}
This is in fact exactly the {\it Ansatz} used by Carter et.al. [9] for the
stationary open string. Furthermore the assumption in this paper of
axially symmetric backgrounds
is by the "duality" transformation (32) turned into the assumption of
stationary
backgrounds of Ref.9, so everything is consistent.

One should however not overestimate the importance of this "duality"
transformation
between a stationary open string and an oscillating closed (circular) string.
A special solution for (say) the open string of Ref.9
in some special "physical" gravitational and electromagnetic background (for
instance a
black hole metric) gives by the "duality" transformation (32) immediately a
special solution for the oscillating circular string in a gravitational and
electromagnetic background with the time coordinate and azimuthal angle
interchanged, which will generally not be a background of any physical
interest.
The "duality" transformation is however useful as long as we only consider the
general formulas describing the strings, for instance the Hamiltonians. An
other
example is related to the electromagnetic properties of the strings. The
electromagnetic spacetime current density is defined by:
\begin{equation}
J^\mu\equiv\frac{\delta}{\delta A_\mu}\left( L(\omega)\sqrt{-\det
G_{\alpha\beta}}\right)=2\frac{dL}{d\omega}\sqrt{-\deg
G_{\alpha\beta}}G^{\alpha\beta}
(\Phi_{,\alpha}+A_\nu x^\nu_{,\alpha})x^\mu_{,\beta}.
\end{equation}
World-sheet charge and current densities $\rho$ and $j$, respectively, can then
be defined as projections onto the world-sheet:
\begin{equation}
J^\mu\equiv j\frac{\partial x^\mu}{\partial\sigma}-\rho\frac{\partial x^\mu}
{\partial\tau},\hspace*{1cm}\frac{\partial\rho}{\partial\tau}=
\frac{\partial j}{\partial\sigma},
\end{equation}
where the continuity equation follows from the equation of motion for $\Phi$.
Using the {\it Ansatz} (5) we find for the circular string:
\begin{equation}
\rho=\Omega,
\end{equation}
\begin{equation}
j=\frac{1}{g_{\phi\phi}}[\Omega g_{a\phi}\frac{dx^a}{d\tau}+
2(n+A_\phi)\sqrt{-\gamma}\frac{dL}{d\omega}].
\end{equation}
In the world-sheet coordinates $(\tau, \sigma)$ the charge density of the
circular string is constant
whereas the current density is given by the complicated $\tau$-dependent
expression (37). Now
performing the "duality" transformation (32) we find that the stationary open
string has a constant current density but a complicated $\sigma$-dependent
charge density, the form of which is given explicitly by equation (37) after
interchanging $(\tau,\sigma)$ and $(t,\phi)$. This is of course in agreement
with the results originally found by Carter et.al. [9].
\vskip 6pt
We conclude this paper with a few remarks on separability and integrability. An
important result of this paper is the reduction of a complicated string problem
to a simpler point particle problem represented by the point particle
Hamiltonians
(19) and (24). It is however well-known that the separation of the
corresponding
equations of motion (for instance by the Hamilton-Jacobi method) is an
intricate
problem, that can usually not be solved unless there are additional symmetries
present.
A special gravitational and electromagnetic background that has attracted a
great deal of interest concerning separability is represented by the
Kerr-Newman
black hole metric and electromagnetic potential (using Boyer-Lindquist
coordinates) [11]:
\begin{equation}
ds^2=-\frac{\Delta}{\rho^2}[dt-a\sin^2\theta
d\phi]^2+\frac{\sin^2\theta}{\rho^2}
[(r^2+a^2)d\phi-adt]^2+\frac{\rho^2}{\Delta}dr^2+\rho^2d\theta^2,
\end{equation}
\begin{equation}
A_\mu=\frac{Qr}{\rho^2}(-1,0,0,a\sin^2\theta),
\end{equation}
where $\Delta=r^2-2Mr+a^2+Q^2$ and $\rho^2=r^2+a^2\cos^2\theta$.

The separation of the simple point particle geodesics of this background was
performed by Carter [12], and this result initiated a comprehensive program
involving spin fields, strings, dyons,... in more general backgrounds also. Of
all these results we shall only mention one of the latest namely the separation
of the equations of motion for a charge-current carrying open stationary string
of the Kaluza-Klein type ($L(\omega)=\sqrt{1+\omega}$) in the Kerr-de Sitter
background [9].

The "duality" transformation (32) relating the open stationary string and the
oscillating circular string might give us hope that similar results can be
obtained for the circular string discussed in this paper, but unfortunately
this does not seem to hold. The potential (28) is for the Kerr-Newman
background (38),(39):
\begin{equation}
V=\frac{1}{2}[((r^2+a^2)^2-a^2\Delta\sin^2\theta)\frac{\sin^2\theta}{\rho^2}
+\Omega^2+(n+\frac{aQr\sin^2\theta}{\rho^2})^2],
\end{equation}
which is not of the separable form of the first term of the Hamiltonian (24),
not even in the spherically symmetric case $a=0$. The point seems to be that
when our circular string is outside the equatorial plane of the black hole, it
experiences non-central forces from the string tension and the electromagnetic
self-interaction, thus reducing the symmetries of the background. For the
Witten model ($L(\omega)=1+\omega/2$) represented by the potential (30) a
similar conclusion can be made.

It should be stressed that this does not necessarily mean that the system is
not integrable. The equations of motion may still be separable in an other
system of coordinates. This and related issues will be discussed in a
forthcoming publication.
\newpage
\begin{centerline}
{\bf References}
\end{centerline}
\begin{enumerate}
\item Nambu Y 1970 Lectures at the Copenhagen Summer Symposium, Goto T 1971
      {\it Prog. Theor. Phys.} {\bf 46} 1560
\item Davidson A and Wali K C 1988 {\it Phys. Lett.} {\bf B213} 439,
      1991 {\it Nucl. Phys.} {\bf B348} 581
\item Balachandran A P, Skagerstam B S and Stern A 1979 {\it Phys. Rev.}
      {\bf D20} 439, Nielsen N K and Olesen P 1987 {\it Nucl. Phys.}
      {\bf B291} 829, 1988 {\it Nucl. Phys.} {\bf B298} 776, Larsen A L
      1991 {\it Phys. Lett.} {\bf B273} 375, 1992 {\it Mod. Phys. Lett.}
      {\bf 7A} 2913, Spergel D N, Piran T and Goodman J 1987
      {\it Nucl. Phys.} {\bf B291} 847
\item Nielsen N K 1980 {\it Nucl. Phys.} {\bf B167} 249
\item Witten E 1985 {\it Nucl. Phys.} {\bf B249} 557, Ostriker J P, Thompson C
      and Witten E 1986 {\it Phys. Lett.} {\bf B180} 231
\item Copeland E, Hindmarsh M and Turok N 1987 {\it Phys. Rev. Lett.}
      {\bf 58} 1910, Vilenkin A and Vachaspati T 1987 {\it Phys. Rev. Lett.}
      {\bf 58} 1041
\item Carter B 1989 {\it Phys. Lett.} {\bf B224} 61
\item Frolov V P, Skarzhinsky V D, Zelnikov A I and Heinrich O 1989
      {\it Phys. Lett.} {\bf B224} 255, Carter B and Frolov V P 1989
      {\it Class. Quantum Grav.} {\bf 6} 569, Carter B 1990 {\it Class.
      Quantum Grav.} {\bf 7} L69
\item Carter B, Frolov V P and Heinrich O 1991 {\it Class. Quantum Grav.}
      {\bf 8} 135
\item Larsen A L 1992 {\it Phys. Lett.} {\bf A170} 174
\item See for instance Misner C W, Thorne K S and Wheeler J A 1973 Gravitation
      ({\it Freeman, San Francisco, CA})
\item Carter B 1967 {\it Phys. Rev.} {\bf 174} 1559
\end{enumerate}
\end{document}